\documentclass[conference]{IEEEtran}
\IEEEoverridecommandlockouts

\usepackage{cite}
\usepackage{amsmath,amssymb,amsfonts}
\usepackage{algorithmic}
\usepackage{graphicx}
\usepackage{textcomp}
\usepackage{xcolor}
\usepackage{booktabs}
\usepackage{multirow}
\usepackage{hyperref}
\usepackage{pifont}
\usepackage{array}
\usepackage{subcaption}

\def\BibTeX{{\rm B\kern-.05em{\sc i\kern-.025em b}\kern-.08em
    T\kern-.1667em\lower.7ex\hbox{E}\kern-.125emX}}
\begin{document}

\title{Metadata-Conditioned Audio Transformers for Adaptive Respiratory Sound Classification}

\author{
  \IEEEauthorblockN{George Kontogiannis, Pantelis Tzamalis, Sotiris Nikoletseas}
  \IEEEauthorblockA{Computer Engineering and Informatics Department\\
  University of Patras, Patras, Greece\\
  g.kontogiannis@ac.upatras.gr, tzamalis@ceid.upatras.gr, nikole@ceid.upatras.gr}
  \thanks{Funded by the Horizon Europe project PHOTOMIX, grant agreement ID: 101297290.}
}




\maketitle

\begin{abstract}
Respiratory sound classification from auscultation recordings is confounded 
by heterogeneous acquisition conditions, including recording device, 
auscultation site, and patient characteristics, that introduce systematic 
variability unrelated to the underlying pathology. Although such metadata 
is often available, existing approaches either ignore it or rely on 
naïve fusion strategies that do not model how distinct metadata sources 
influence different components of the acoustic representation.

We propose a family of metadata-conditioning mechanisms for the Audio 
Spectrogram Transformer (AST), progressing from gated residual fusion 
to Feature-wise Linear Modulation (FiLM) adapted to the Transformer 
architecture. We further introduce two novel variants: \emph{Token-Aware 
FiLM (TAFiLM)}, which applies distinct modulation to classification and 
patch tokens, and \emph{Soft-Factorized FiLM (SoftFiLM)}, which learns 
soft masks to align metadata factors with disentangled feature subspaces.

Experiments on the ICBHI benchmark demonstrate consistent gains across 
all conditioning strategies. SoftFiLM achieves 64.11\% on the official 
4-class task (+3.94\% over the baseline) and 72.40\% on the 2-class task, 
establishing a new state-of-the-art on the 2-class task and competitive performance on the 4-class benchmark. 

\end{abstract}


\section{Introduction}
\label{sec:introduction}

Respiratory diseases remain a leading cause of morbidity and mortality worldwide, with auscultation serving as a fundamental diagnostic tool for identifying pulmonary abnormalities~\cite{iotlabaudio}. 
The advent of digital stethoscopes has enabled the collection of large-scale respiratory sound databases, paving the way for automated classification systems that could assist clinicians in resource-limited settings. 
The ICBHI 2017 Respiratory Sound Database~\cite{rocha2017alpha} has emerged as the primary benchmark for this task.

Deep learning approaches have achieved substantial progress on this benchmark under the official 60-40\% train--test split. 
Early CNN-based methods applied ResNet~\cite{gairola2021respirenet} and VGG architectures to mel-spectrogram representations, while hybrid CNN-RNN networks addressed temporal modeling of respiratory phases~\cite{perna2019deep, ma2019lungbrn, nguyen2022lung}.  
More recently, Transformer-based architectures have set new state-of-the-art results: the Audio Spectrogram Transformer (AST)~\cite{gong2021ast} and its variants have demonstrated strong performance through transfer learning from large-scale audio pretraining~\cite{bae2023patch, moummad2023pretraining}, while more recent approaches \cite{kim2024bts, toikkanen2025improving}, use multimodal audio-text models to achieve higher performances.

Despite these advances, a critical limitation persists: \textit{most methods treat the acoustic signal in isolation}, ignoring the available metadata that accompanies each recording. 
The ICBHI dataset exhibits substantial \textit{acquisition heterogeneity} arising from multiple sources: four different recording devices with distinct frequency responses and noise characteristics; seven anatomical auscultation sites spanning anterior, posterior, lateral chest, and tracheal positions; and diverse patient demographics including sex, age, BMI, and breathing cycle duration. 
This heterogeneity introduces acquisition-induced distributional bias that confound the underlying pathological signatures.
Prior work~\cite{gairola2021respirenet} demonstrated that device distribution is highly skewed and classification performance correlates strongly with device-specific training set size, yet proposed only post-hoc device-specific fine-tuning rather than explicit conditioning.
Our statistical analysis of the dataset (Section~\ref{sec:dataset_analysis}) confirms these findings and further reveals significant effects of auscultation site and patient characteristics on classification difficulty.

Only a handful of recent works have attempted to leverage metadata for respiratory sound classification. 
The BTS framework~\cite{kim2024bts} converts metadata into free-text descriptions (e.g., ``recorded with Littmann stethoscope at the anterior left chest'') and fuses text embeddings with audio features via a CLAP-style architecture, achieving modest improvements (+1.17\% Score over audio-only baseline). 
Contrastive learning approaches have used metadata for auxiliary objectives: sex and age were employed as pretext tasks for self-supervised pretraining~\cite{moummad2023pretraining}, while stethoscope types were treated as separate domains for domain-adversarial training~\cite{kim2024stethoscope} or domain generalization~\cite{kim2025adaptive}. 

However, these methods share common limitations: they either require expensive multimodal encoders, operate only at the representation learning stage without conditioning inference, address device and demographic factors in isolation, or employ naive fusion strategies (e.g., feature concatenation) that fail to capture how different metadata sources affect distinct aspects of the acoustic representation. 
Crucially, \textit{no prior work applies feature-wise conditioning mechanisms} that can dynamically modulate the acoustic encoder's internal representations based on the full recording context.

Feature-wise Linear Modulation (FiLM)~\cite{perez2018film} offers an elegant solution to this gap. 
Originally proposed for visual question answering, FiLM enables a conditioning signal to modulate neural network activations through learned scale ($\gamma$) and shift ($\beta$) parameters applied after normalization layers. 
However, while FiLM was originally designed for CNNs, its adaptation to Vision Transformers and Audio Spectrogram Transformers remains underexplored, and it has never been applied to respiratory sound classification or medical audio analysis more broadly.


Standard FiLM applies uniform modulation across all spatial positions (or Transformer tokens) and treats the conditioning signal as a single vector, which overlooks two key points. First, in Transformers, the classification token (CLS) aggregates global information while patch tokens capture local spectro-temporal patterns, yet standard FiLM modulates them identically. Second, different metadata sources (device, site, demographics) affect distinct aspects of the acoustic signal: device shapes spectral characteristics and noise, site determines audible lung sounds, and patient factors influence breath intensity and baseline. Uniform treatment of all metadata ignores this factor-specific structure.

In this paper, we formalize metadata conditioning as structured feature modulation within Transformer-based acoustic encoders. Rather than treating metadata as auxiliary inputs, we model acquisition context as a factorized conditioning signal that reshapes internal representations throughout the network hierarchy. Our main contributions are:  
\begin{enumerate}
    \item We perform the first \textbf{comprehensive metadata-level analysis} of ICBHI, quantifying device bias, auscultation site effects, and demographic confounds. This reveals systematic domain shifts, motivates \textbf{metadata conditioning}, and identifies discriminative versus confounding factors.  
    \item We provide the first \textbf{systematic study of metadata conditioning} for respiratory sound classification, adapting \textbf{Feature-wise Linear Modulation (FiLM)} to the \textbf{Audio Spectrogram Transformer} and demonstrating its effectiveness for encoding acquisition context.  
    \item We introduce \textbf{Token-Aware FiLM (TAFiLM)} and \textbf{Soft-Factorized FiLM (SoftFiLM)}. TAFiLM modulates \textbf{classification vs. patch tokens} separately, while SoftFiLM learns \textbf{factor-aligned soft masks} partitioning features into metadata-specific subspaces.  
    \item We achieve \textbf{competitive four-class} and \textbf{state-of-the-art two-class} ICBHI results: SoftFiLM reaches \textbf{64.11\% (+3.94\%)} and \textbf{72.40\% (+2.32\%)} respectively. Ablations show \textbf{full fine-tuning} and conditioning all layers maximize gains, and qualitative analysis confirms interpretable factor disentanglement. Code is released for reproducibility.
    \footnote{Code is available at: \url{https://github.com/kontogiannisg/Metadata-Conditioned_Audio_Transformers}}

\end{enumerate}


\section{ICBHI Dataset and Metadata Analysis}
\label{sec:dataset_analysis}
We use the ICBHI Respiratory Sound Database~\cite{rocha2017alpha}, which contains lung sound recordings from 126 patients (920 sessions) collected using four different devices—\textit{Meditron}, \textit{LittC2SE}, \textit{Litt3200}, and \textit{AKGC417L}—across seven auscultation sites: \textit{anterior left (Al)}, \textit{anterior right (Ar)}, \textit{posterior left (Pl)}, \textit{posterior right (Pr)}, \textit{lateral left (Ll)}, \textit{lateral right (Lr)}, and \textit{trachea (Tc)}.  
Each recording includes at least one complete breathing cycle and is annotated for the presence of \textit{crackles}, \textit{wheezes}, or \textit{both} (co-occurring crackle and wheeze). After segmentation, this resulted in a total of 6,898 breathing cycles, each assigned to one of four classes: \textit{Normal} (52.8\%), \textit{Crackle} (27.0\%), \textit{Wheeze} (12.8\%), or \textit{Both} (7.3\%).

We analyze the dataset to quantify metadata-driven
confounds that motivate our conditioning approach. All statistical tests use
\textit{Generalized Estimating Equation (GEE)} models clustered by patient to account for repeated measures.

\noindent\textbf{Device bias.}
The dataset is dominated by a single device: AKGC417L contributes 63\%
of all cycles and over 80\% of \textit{Crackle} labels in both splits. One
stethoscope (LittC2SE, 594 cycles) appears exclusively in training, creating a
complete domain gap at test time. Since 122 of 126 patients used only one
device, patient identity and device are largely confounded. Therefore, a model can learn
device-specific spectral signatures as a proxy for pathology.

\noindent\textbf{Site effects.}
Posterior and lateral sites contain 35--40\% abnormal cycles, while tracheal
recordings are 71\% Normal with zero \textit{Both} events in the test split.
GEE analysis confirms that posterior sites have 2--4$\times$ higher odds of
\textit{Crackle} than tracheal ($p_{\mathrm{bonf}}{<}0.01$), and the right
posterior site shows $\approx$4.2$\times$ odds for \textit{Both}. Site
distribution also shifts between splits, introducing additional domain
mismatch.

\noindent\textbf{Continuous covariates.}
Sex was not significantly associated with any sound class after correction.
Abnormal cycles are 0.1--0.4\,s longer than Normal ones ($p{<}0.05$, mixed-effects
model). Age is associated with higher odds of \textit{Both}
(OR$\,{=}\,$1.07, $p_{\mathrm{bonf}}{=}$0.024), while higher BMI is associated
with more \textit{Normal} cycles (OR$\,{=}\,$1.07, $p_{\mathrm{bonf}}{=}$0.038)
and fewer \textit{Both} (OR$\,{=}\,$0.93, $p_{\mathrm{bonf}}{=}$0.032).
Sex was therefore excluded from the conditioning vector, which consists of
age, BMI, and cycle duration.

These findings establish that device, site, and continuous covariates each
introduce systematic bias into the acoustic distribution. This motivates
explicit metadata conditioning rather than treating recordings as i.i.d.\
audio samples.

\section{Methods}
\label{sec:methods}

We present a family of metadata conditioning approaches for respiratory sound
classification, building from uniform modulation to factor-aligned conditioning.
All methods share the Audio Spectrogram Transformer (AST)~\cite{gong2021ast}
as the acoustic backbone.

\subsection{Baseline and Metadata Representation}
\label{sec:baseline}

We use AST-Base~\cite{gong2021ast} pretrained on ImageNet and AudioSet as the
acoustic encoder. Log-mel spectrograms are split into $16{\times}16$ patches,
linearly projected to $D$ dimensions, prepended with \texttt{[CLS]} and
\texttt{[DIST]} tokens~\cite{touvron2021training}, and processed by $L$
Transformer layers. The final representation averages both special tokens:
$h_{\mathrm{audio}} = (Z_L[\texttt{CLS}] + Z_L[\texttt{DIST}])/2$.

For metadata, device index $d$ and site index $s$ are mapped to learned
embeddings $e_{\mathrm{dev}}$ and $e_{\mathrm{site}}$; age, BMI, and duration
form a standardized continuous vector $m_{\mathrm{cont}} \in \mathbb{R}^3$.

\subsection{Gated Residual Fusion (GRF)}
\label{sec:gated_fusion}

As a simple baseline, we project the concatenated metadata
$m = [e_{\mathrm{dev}};\, e_{\mathrm{site}};\, m_{\mathrm{cont}}]$ to match
the audio embedding dimension via an MLP and integrate it through a gated
residual: $\tilde{h} = h_{\mathrm{audio}} + \alpha \cdot \mathrm{MLP}(m)$,
where $\alpha$ is a learnable scalar. This late-fusion approach can only
modify the final representation; metadata cannot influence intermediate
feature extraction.

\subsection{Feature-wise Linear Modulation (FiLM)}
\label{sec:film}

FiLM~\cite{perez2018film} enables metadata to modulate intermediate
representations. A shared encoder produces a conditioning embedding
$h_m = g(m)$, from which layer-specific scale and shift parameters are
generated:
\begin{align}
    \gamma_\ell &= \mathbf{1} + W_\ell^{(\gamma)} h_m, \quad
    \beta_\ell = W_\ell^{(\beta)} h_m.
\end{align}
For conditioned layers $\ell \in \mathcal{B}$, modulation is applied after
layer normalization and before the FFN:
\begin{align}
    \hat{Z}_\ell &= \gamma_\ell \odot \mathrm{LN}_2(\tilde{Z}_\ell)
                    + \beta_\ell,
\end{align}
where $\odot$ denotes element-wise multiplication broadcast across all tokens.
The identity-centered parameterization ensures near-identity modulation at
initialization, preserving pretrained features.

FiLM modulation is applied after the second LayerNorm (pre-FFN normalization) and before the feed-forward sublayer, preserving the residual structure of the Transformer block.

\subsection{Token-Aware FiLM (TAFiLM)}
\label{sec:tafilm}

Standard FiLM applies identical parameters to all tokens, ignoring the
distinct roles of classification and patch tokens. TAFiLM generates separate
modulation for each:
\begin{align}
    \hat{Z}_\ell[t,:] = \begin{cases}
        \gamma_\ell^{\mathrm{cls}} \odot \bar{Z}_\ell[t,:]
            + \beta_\ell^{\mathrm{cls}},
            & t \in \{\texttt{CLS}, \texttt{DIST}\}, \\[3pt]
        \gamma_\ell^{\mathrm{patch}} \odot \bar{Z}_\ell[t,:]
            + \beta_\ell^{\mathrm{patch}},
            & \text{otherwise},
    \end{cases}
\end{align}
where $(\gamma^{\mathrm{cls}}, \beta^{\mathrm{cls}})$ and
$(\gamma^{\mathrm{patch}}, \beta^{\mathrm{patch}})$ are generated by
separate linear projections from $h_m$. This lets metadata adjust global
aggregation (CLS) and local spectro-temporal extraction (patches) independently,
at the cost of doubling FiLM parameters per layer ($<$0.4M additional).

\subsection{Soft-Factorized FiLM (SoftFiLM)}
\label{sec:softfilm}

FiLM and TAFiLM treat all metadata sources uniformly through a combined
embedding. SoftFiLM instead processes each source through a dedicated encoder
and learns soft masks that determine which feature dimensions each source
should modulate.

\noindent\textbf{Factor-specific encoding.}
Each metadata source feeds a separate encoder and FiLM generator:
\begin{align}
    h^{(k)} &= g^{(k)}(\cdot), \quad
    \gamma_\ell^{(k)}, \beta_\ell^{(k)} = f_\ell^{(k)}(h^{(k)}), \notag \\
    & \qquad k \in \{\mathrm{dev}, \mathrm{site}, \mathrm{cont}\}.
\end{align}
Generators are initialized with small Xavier weights~\cite{glorot2010} (gain$\,{=}\,0.1$) to
ensure non-zero gradients reach the mask parameters from the first step.

\noindent\textbf{Learnable soft masks.}
For each source $k$, a learnable logit vector $\omega^{(k)} \in \mathbb{R}^D$
produces mask weights via temperature-scaled sigmoid:
$w^{(k)} = \sigma(\omega^{(k)}/\tau)$.
Temperature $\tau$ is annealed from 1.0 to 0.2 via cosine schedule, yielding
soft masks early in training and near-binary factorization at convergence.
Mask logits use $10{\times}$ higher learning rate and are initialized to
partition dimensions into approximate thirds with scale $s{=}0.5$.

\noindent\textbf{Mask-weighted modulation.}
Combined FiLM parameters are a mask-weighted sum:
\begin{align}
\gamma_\ell 
&= \mathbf{1} 
   + \sum_{k} w^{(k)} \odot
     \big(\gamma_\ell^{(k)} - \mathbf{1}\big), \notag \\
\beta_\ell 
&= \sum_{k} w^{(k)} \odot \beta_\ell^{(k)}.
\end{align}

\noindent\textbf{Regularization.}
An overlap penalty discourages multiple masks from claiming the same
dimension, and a coverage penalty ensures all dimensions are used:
\begin{align}
\mathcal{L}
&= \mathcal{L}_{\mathrm{task}}
 + \lambda_{\mathrm{ov}}
   \sum_{k \neq k'} \frac{1}{D}
   \sum_c w_c^{(k)} w_c^{(k')} \notag \\
&\quad
 + \lambda_{\mathrm{cov}} \frac{1}{D}
   \sum_c \left(1 - \max_k w_c^{(k)}\right).
\end{align}


Unlike naïve multi-branch FiLM, which sums independent modulations uniformly, SoftFiLM uses learnable feature-axis masks to assign disjoint subspaces of the Transformer embedding to different metadata sources. Joint optimization with overlap and coverage regularization enforces structured factor alignment, producing an \textit{interpretable, metadata-partitioned feature space} rather than unstructured additive modulation.

\begin{table*}[h!]
\centering
\caption{Comparison with state-of-the-art methods on the ICBHI lung sound classification task using the official 60--40\% train--test split. 
IN, AS, and LA refer to ImageNet, AudioSet, LAION-Audio-630K, respectively. 
$S_p$: Specificity, $S_e$: Sensitivity, Score $= (S_p + S_e)/2$.
The \textbf{best} and \underline{second best} results are highlighted. Scores marked with $*$ denote the previous state-of-the-art performance.}
\label{tab:sota_comparison}

\resizebox{\linewidth}{!}{%
\begin{tabular}{lllllccc}
\toprule
\textbf{Method} & \textbf{Backbone} & \textbf{Pretrain} & \textbf{Venue} & \textbf{Meta} & $\boldsymbol{S_p}$ (\%) & $\boldsymbol{S_e}$ (\%) & \textbf{Score} (\%) \\
\midrule
\multicolumn{8}{c}{\textit{\textbf{4-class evaluation}}} \\
\midrule
Kim \textit{et al.} (SG-SCL) \cite{kim2024stethoscope}       & AST              & IN + AS & \textit{ICASSP'24}        & \ding{51} & 79.87 & 43.55 & 61.71 \\
Kim \textit{et al.} (RepAugment) \cite{kim2024repaugment}& AST            & IN + AS & \textit{EMBC'24}          & \ding{55} & 82.47 & 40.55 & 61.51 \\
Daisuke \textit{et al.} (M2D) \cite{niizumi2024masked}        & M2D ViT          & AS      & \textit{TASLP'24}         & \ding{55} & 81.51 & 45.08 & 63.29 \\
Kim \textit{et al.} (BTS) \cite{kim2024bts}            & CLAP             & LA      & \textit{INTERSPEECH'24}   & \ding{51} & 81.40 & 45.67 & 63.54 \\
Wang \textit{et al.} (HTS-AT) \cite{wang2024lightweight}    & HTS-AT           & IN + AS & \textit{ICSIP'24}         & \ding{55} & 79.61 & 48.77 & 64.19 \\
MVST \cite{he2024multi}                                 & AST              & IN + AS & \textit{ICASSP'24}        & \ding{55} & 81.99 & \underline{51.10} & \underline{66.55} \\
LungAdapter \cite{xiao2024lungadapter}                   & AST             & IN + AS & \textit{INTERSPEECH'24}   & \ding{55} & 80.43 & 44.37 & 62.40 \\
CycleGuardian \cite{chu2025cycleguardian}               & CNN               & --      & \textit{CIS'25}           & \ding{55} & 82.06 & 44.47 & 63.26 \\
Fraihi \textit{et al.} (Patch-Mix CL + FBS) \cite{fraihi2025improving} & AST   & IN + AS & \textit{arXiv'25}         & \ding{51} & 84.35 & 43.67 & 64.01 \\
Wei \textit{et al.} (DDE-MAE) \cite{wei2025disentangling}   & Audio-MAE        & AS      & \textit{INTERSPEECH'25}   & \ding{55} & 69.32 & \textbf{53.69} & 61.50 \\
Toikkanen \textit{et al.} (BTS-d[$k$=15]) \cite{toikkanen2025improving}  & CLAP  & LA  & \textit{INTERSPEECH'25}  & \ding{51} & 82.89 & 45.90 & 64.39 \\
Dong \textit{et al.} (AFF + DDL + $L_{BD}$) \cite{dong2025adaptive}       & AST   & IN + AS & \textit{INTERSPEECH'25}   & \ding{55} & 85.13 & 45.94 & 65.53 \\
Dong \textit{et al.} (multi-view) \cite{fusion}             & AST+Time-Domain   & IN + AS & \textit{BSPC'25}          & \ding{55} & \textbf{85.99} & 49.11 & \textbf{67.55}$^*$ \\
\midrule
\textbf{GRF-AST [ours]}       & AST & IN + AS & \textit{EUSIPCO'26} & \ding{51} & 74.47$_{\scriptstyle \pm 1.10}$ & 47.91$_{\scriptstyle \pm 0.85}$ & 61.19$_{\scriptstyle \pm 0.90}$ \\
\textbf{FiLM-AST [ours]}      & AST & IN + AS & \textit{EUSIPCO'26} & \ding{51} & 77.26$_{\scriptstyle \pm 1.25}$ & 47.49$_{\scriptstyle \pm 0.95}$ & 62.37$_{\scriptstyle \pm 1.05}$ \\
\textbf{TAFiLM-AST [ours]}    & AST & IN + AS & \textit{EUSIPCO'26} & \ding{51} & 78.84$_{\scriptstyle \pm 1.30}$ & 48.20$_{\scriptstyle \pm 1.10}$ & 63.52$_{\scriptstyle \pm 1.20}$ \\
\textbf{SoftFiLM-AST [ours]}  & AST & IN + AS & \textit{EUSIPCO'26} & \ding{51} & 78.92$_{\scriptstyle \pm 1.40}$ & 49.30$_{\scriptstyle \pm 1.25}$ & 64.11$_{\scriptstyle \pm 1.15}$ \\

\midrule
\multicolumn{8}{c}{\textit{\textbf{2-class evaluation}}} \\
\midrule
Nguyen \textit{et al.} (CoTuning) \cite{nguyen2022lung}   & ResNet50     & IN      & \textit{TBME'22}          & \ding{55} & 79.34 & 50.14 & 64.74 \\
Bae \textit{et al.} (Patch-Mix CL) \cite{patchmix}  & AST          & IN + AS & \textit{INTERSPEECH'23}   & \ding{55} & 81.66 & 55.77 & 68.71 \\
SG-SCL \cite{kim2024stethoscope}                         & AST          & IN + AS & \textit{ICASSP'24}        & \ding{51} & 79.87 & 57.97 & 68.93 \\
Fraihi \textit{et al.} (Patch-Mix CL + FBS) \cite{fraihi2025improving} & AST  & IN + AS & \textit{arXiv'25}         & \ding{51} & 75.17 & 65.0 & $70.08^*$ \\ 
\midrule

\textbf{GRF-AST [ours]}       & AST & IN + AS & \textit{EUSIPCO'26} & \ding{51} & 79.10$_{\scriptstyle \pm 1.20}$ & 61.25$_{\scriptstyle \pm 0.90}$ & 70.17$_{\scriptstyle \pm 0.74}$ \\
\textbf{FiLM-AST [ours]}      & AST & IN + AS & \textit{EUSIPCO'26} & \ding{51} & 75.49$_{\scriptstyle \pm 1.10}$ & \underline{67.45$_{\scriptstyle \pm 1.30}$} & \underline{71.47$_{\scriptstyle \pm 1.00}$} \\
\textbf{TAFiLM-AST [ours]}    & AST & IN + AS & \textit{EUSIPCO'26} & \ding{51} & 77.80$_{\scriptstyle \pm 1.35}$ & 66.40$_{\scriptstyle \pm 1.25}$ & 71.10$_{\scriptstyle \pm 0.95}$ \\
\textbf{SoftFiLM-AST [ours]}  & AST & IN + AS & \textit{EUSIPCO'26} & \ding{51} & 77.20$_{\scriptstyle \pm 1.40}$ & \textbf{67.60$_{\scriptstyle \pm 1.15}$} & \textbf{72.40$_{\scriptstyle \pm 0.80}$} \\
\bottomrule
\end{tabular}%
}
\end{table*}

\section{Experiments}
\label{sec:experiments}

\subsection{Experimental Setup}
\label{sec:setup}

We evaluate on the official split of the ICBHI dataset (train-test split as 60-40\%). Audio is resampled to 16\,kHz,
fixed to 10\,s via cross-fade repetition, and converted to $128{\times}1024$
log-mel spectrograms normalized with AudioSet statistics~\cite{gong2021ast}.
We apply SpecAugment~\cite{park2019specaugment}, additive noise, speed
perturbation, and pitch shifting. All methods use AST-Base~\cite{gong2021ast}
(87M parameters, ImageNet + AudioSet pretrained) with a two-layer MLP head
(hidden dim 64, dropout 0.4). We formulate the task as multi-label classification with two sigmoid outputs 
(crackle, wheeze) and optimize binary cross-entropy (BCE) loss. Conditioning modules add $<$1\% parameters.
Training runs for 100 epochs (batch 16, AdamW, cosine schedule with 10-epoch
warmup, base LR $3{\times}10^{-5}$) with differential learning rates:
$0.01{\times}$ backbone, $0.5{\times}$ embeddings, $2{\times}$ classifier.
SoftFiLM adds mask overlap regularization ($\lambda{=}0.01$). We report
Sensitivity ($S_e$), Specificity ($S_p$), and ICBHI Score
$({S_p + S_e})/{2}$~\cite{rocha2017alpha}, averaged over five runs.

\subsection{Results}
\label{sec:results}

Table~\ref{tab:sota_comparison} compares our methods against prior work.
On the 4-class task, each variant improves over the baseline AST (60.17\%):
FiLM 62.37\%, TAFiLM 63.52\%, SoftFiLM 64.11\% (+3.94\%), confirming that
more structured conditioning progressively better captures how metadata shapes
acoustic representations; SoftFiLM also surpasses metadata-aware
SG-SCL~\cite{kim2024stethoscope} (61.71\%) and BTS~\cite{kim2024bts}
(63.54\%) with a simpler mechanism. On the 2-class task, SoftFiLM reaches
72.40\% (+2.32\% over the previous best~\cite{fraihi2025improving}),
consistent with our dataset analysis (Section~\ref{sec:dataset_analysis}):
device and site biases primarily affect normal-vs-abnormal discrimination,
where metadata conditioning is most effective. The leading 4-class-only
methods (MVST~\cite{he2024multi}: 66.55\%, Dong et al.~\cite{fusion}:
67.55\%) do not report 2-class results; among those that
do---SG-SCL (61.71\%/68.93\%), Patch-Mix CL (62.37\%/68.71\%), Fraihi
et al.\ (64.01\%/70.08\%)---SoftFiLM achieves the highest 2-class score.




\vspace{-4pt}
\subsection{Ablation: Conditioning Layer Selection}
\label{sec:ablation}
Table~\ref{tab:ablation_conditioning} shows the effect of which Transformer
layers receive FiLM conditioning, with full backbone fine-tuning.
Conditioning only layers $\{10, 11\}$ gives modest gains (e.g., SoftFiLM:
62.45\%), while conditioning all 12 layers maximizes performance (64.11\%). This suggests that that metadata-driven modulation benefits both early layers, where
device-specific spectral normalization may help, and later layers where
task-specific decisions are refined. All main results use all-layer
conditioning.

\begin{table}[h!]
\centering
\caption{Effect of conditioning layer selection on ICBHI Score (\%, 4-class).
Full fine-tuning. Best per method in \textbf{bold}.}
\label{tab:ablation_conditioning}
\begin{tabular}{lccc}
\toprule
\textbf{Method} & Last 2 & Last 6 & All 12 \\
\midrule
Baseline AST & \multicolumn{3}{c}{60.17 (no conditioning)} \\
\midrule
FiLM     & 61.02 & 61.68 & \textbf{62.37} \\
TAFiLM   & 61.89 & 62.74 & \textbf{63.52} \\
SoftFiLM & 62.45 & 63.31 & \textbf{64.11} \\
\bottomrule
\end{tabular}
\end{table}

\vspace{-9pt}
\subsection{Qualitative Analysis}
\label{sec:mask_analysis}
\begin{figure}[t]
    \centering
    
    \includegraphics[width=\columnwidth]{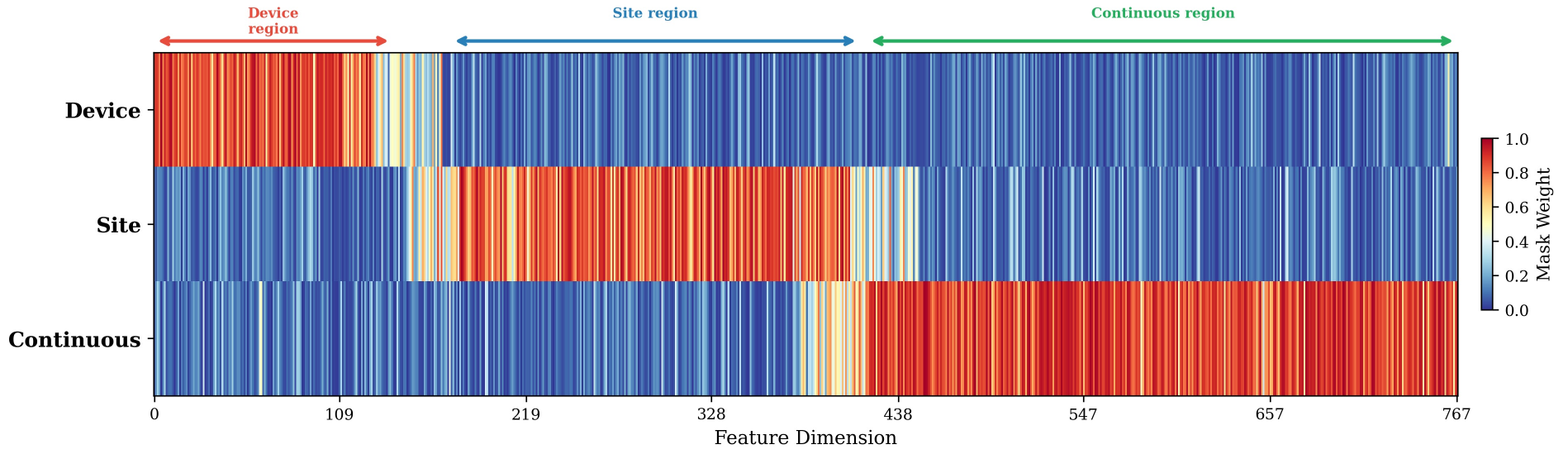}
    
    \vspace{2mm}
    
    \begin{subfigure}{0.56\columnwidth}
        \centering
        \includegraphics[width=\linewidth]{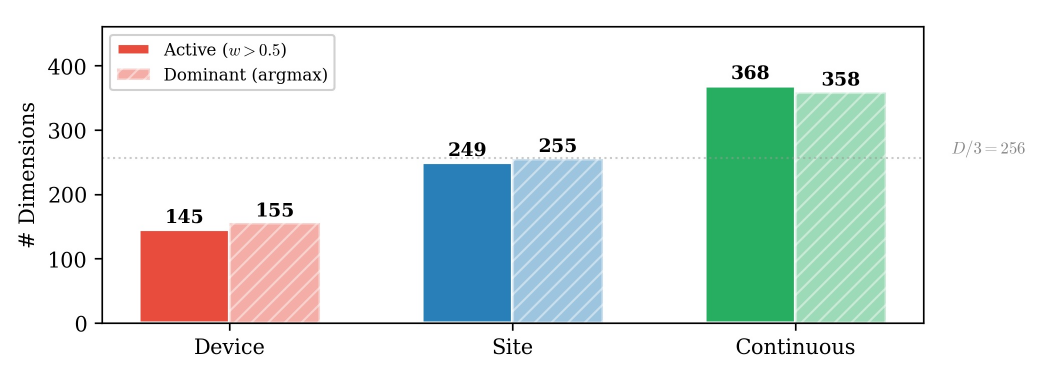}
    \end{subfigure}
    \hfill
    \begin{subfigure}{0.28\columnwidth}
        \centering
        \includegraphics[width=\linewidth]{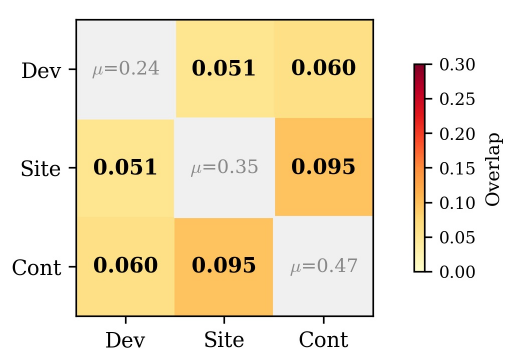}
    \end{subfigure}

    \caption{Learned SoftFiLM mask structure ($D{=}768$, $\tau{=}0.25$).
    Top: mask activations across dimensions. Bottom-left: dimensions
    claimed per factor. Bottom-right: pairwise overlap.}
    \vspace{-15pt}
    \label{fig:mask_analysis}
\end{figure}

Figure~\ref{fig:mask_analysis} shows the learned masks after training.
Although initialized to equal thirds, the model reallocates dimensions
unevenly: device (low-cardinality) claims the fewest; continuous patient
metadata claims nearly half. The factors occupy largely disjoint regions
of the feature space, with 95.6\% coverage and pairwise overlap below 0.1.
The slightly higher site--continuous overlap (0.095) is consistent with
recording site correlating with body characteristics that also appear in
the continuous metadata. This anisotropic allocation supports the hypothesis
that metadata influences the embedding space along structured, low-dimensional
axes rather than uniformly across features.

\vspace{-5pt}
\section{Conclusion}
\label{sec:conclusion}
We proposed three FiLM-based conditioning variants for respiratory sound
classification that explicitly model acquisition and patient-dependent variability in the
ICBHI dataset. SoftFiLM, which learns soft masks assigning feature
dimensions to metadata sources, achieved ICBHI Scores of 72.40\% (2-class)
and 64.11\% (4-class). Its learned masks reveal an interpretable
factorization: fewer dimensions for device identity, more for continuous
patient metadata. Future work could combine token-axis (TAFiLM) and feature-axis (SoftFiLM) partitioning, and extend the approach toward improved robustness to unseen metadata distributions, such as previously unseen devices or acquisition conditions.

\bibliographystyle{IEEEtran}
\bibliography{references}

\end{document}